\documentclass[
reprint,
superscriptaddress,
amsmath,amssymb,
prl
]{revtex4-1}
\usepackage{xcolor}
\usepackage{color}
\usepackage[normalem]{ulem}

\usepackage{graphicx}
\usepackage{subfigure}
\usepackage{dcolumn}
\usepackage{bm}
\usepackage{xcolor}%
\usepackage{amsmath}
\usepackage{verbatim}
\usepackage{mathtools}
\usepackage{epstopdf}
\usepackage{float}
\usepackage{multirow}

\begin{document}
	
	\title{Friction-controlled entropy-stability competition in granular systems}
	
	\author{Xulai Sun}
	\affiliation{School of Physics and Astronomy, Shanghai Jiao Tong University, Shanghai 200240, China}
	
	\author{Walter Kob}
	\affiliation{School of Physics and Astronomy, Shanghai Jiao Tong University, Shanghai 200240, China}
	\affiliation{Laboratoire Charles Coulomb, University of Montpellier and CNRS, F34095 Montpellier, France}
	
	\author{Raphael Blumenfeld}
	\affiliation{Gonville \& Caius College and Cavendish Laboratory, University of Cambridge, Cambridge CB2 1TA, UK}
	
	\author{Hua Tong}
	\affiliation{School of Physics and Astronomy, Shanghai Jiao Tong University, Shanghai 200240, China}
	
	\author{Yujie Wang}
	\affiliation{School of Physics and Astronomy, Shanghai Jiao Tong University, Shanghai 200240, China}
	
	\author{Jie Zhang}
	\email{jiezhang2012@sjtu.edu.cn}
	\affiliation{School of Physics and Astronomy, Shanghai Jiao Tong University, Shanghai 200240, China}
	\affiliation{Institute of Natural Sciences, Shanghai Jiao Tong University, Shanghai 200240, China}
	
	\date{\today}
	
	\begin{abstract}
		Using cyclic shear to drive a two dimensional granular system, we determine the structural characteristics for different inter-particle friction coefficients. These characteristics are the result of a competition between mechanical stability and entropy, with the latter's effect increasing with friction. We show that a parameter-free maximum-entropy argument alone predicts an exponential cell order distribution, with excellent agreement with the experimental observation. We show that friction only tunes the mean cell order and, consequently, the exponential decay rate and the packing fraction. We further show that cells, which can be very large in such systems, are short-lived, implying that our systems are liquid-like rather than glassy.
		
	\end{abstract}
	
	\maketitle
	
	Dense granular materials show a highly complex response when subjected to repeated cycles of shear, mainly due to the strongly
	dissipative, hysteretic, and nonlinear interactions at the frictional contacts. The structures of such systems self-organize dynamically and show characteristics that on large scales appear to be universal \cite{RafiTakashi2014}. Previous works focused on the motion on the particle scale~\cite{kou2017granular, MailmanPRL} and phase behavior~\cite{PreciselyCyclic,ShearNucleation,ZhangGranMatt}, the slow relaxation of stress and density~\cite{RenJiePRL,PoliquenCompactionEPL,PouliquenCompactionPRL}, and the complex spatio-temporal dynamics~\cite{NagelMemory,DauchotHeterogeneity,LosertReverse}. These dynamics are relevant in a wide range of fields, including the aging and memory of glasses~\cite{SastryRMP}, fatigue of materials~\cite{suresh1998fatigue}, catastrophic collapse of soils and sand in civil and geotechnical engineering~\cite{terzaghi1996soil},
	as well as in geological processes, such as earthquakes and landslides~\cite{scholz2019mechanics}. Despite this multitude of investigations, the nature and role of the contact network, a key quantity of granular systems, are far from being understood and in particular the relation between the properties of this network and the friction between the particles is currently not known. Intuitively, one expects higher friction to give rise to looser structures and, therefore, to overall larger cells in the contact network. 
	It has been proposed that the characteristics of granular packs can be understood from entropic considerations~\cite{Edwards1}. In turn, the entropy should depend on the driving process and friction. To quantify this relation, we have created a set-up that allows to analyse entropy of contact networks for widely different friction coefficients.
	
	\begin{figure}[htb]
		\includegraphics{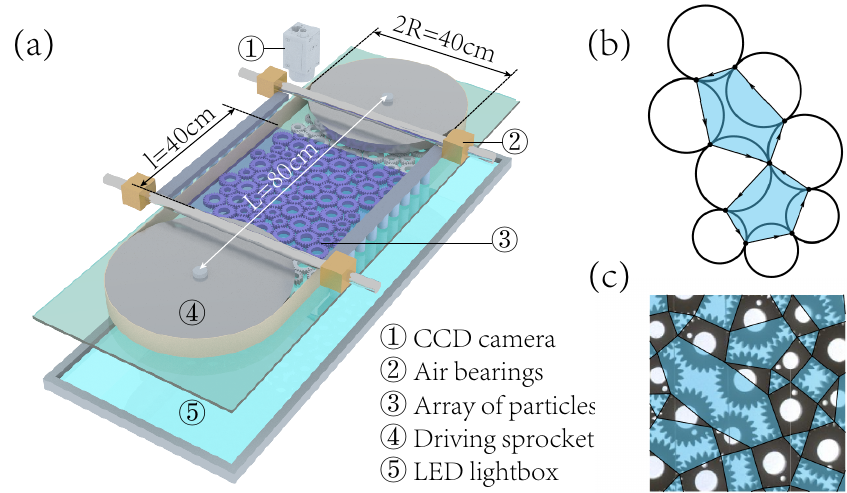}
		\caption{\label{fig1}
			(a) The stadium shear device is mounted on a horizontal glass plate. Particles occupy the region within the stadium, including areas under the sprockets. A constant boundary confining pressure is maintained by two aluminium bars, mounted on four air bearings with constant forces pushing against the lateral walls of the belt. The imaging system includes a CCD camera and a LED panel above and below the device, respectively. 
			(b) A cell is defined as the smallest loop of contacts. Two cells are drawn in the diagram. 
			(c) A part of the gear system, containing cells, shaded blue.}.
	\end{figure}
	
	We consider a two-dimensional (2D) system and focus on its cells, defined as the smallest (aka irreducible) closed loops of contacts between particles that touch at least two neighbors.
	Other than their importance for the mechanical properties of the system, the properties of cells affect heat conduction in granular assemblies and, in three dimensions, the permeability to fluid flow.
	Cells are closely related to quadrons~\cite{Rafi2003,BlumenfeldEdwards06,RafiTakashi2014, RafiTakashi2017}, which are the smallest volume elements of granular assemblies and play a fundamental role in granular statistical mechanics~\cite{Rafi2003, Edwards1, Edwards2,EdwardsReview}.  
	During cyclic shear, contacts are continually broken and made, leading to cells being created and annihilated. 
	Here, we investigate the effect of friction on the structure and quasi-static dynamics.
	Our findings are the following: 1.~The cell order distributions are exponential and can be derived from a maximum entropy argument. 2.~The structural characteristics result from a competition between entropy and mechanical stability, which the friction tunes. Specifically, mechanical stability curbs entropy by increasingly limiting the number of possible cell configurations as the cell order increases, making entropy more dominant as the inter-particle friction increases. 3.~Our systems are liquid-like rather than glassy.
	
	The experimental setup, known as the \emph{stadium shear device}~\cite{Itai}, is sketched in Fig.~\ref{fig1}(a). We used a stepping motor to drive periodically two stainless steel sprockets, connected to each other by a rubber belt which was corrugated on the inside to ensure no-slip between it and the  particles. The particle assembly was sheared between the two parallel sections of the belt in the central region of the device and recirculate under the two sprockets. We observed the particles within the blue shaded area, shown in Fig.~\ref{fig1}(a), which thus mimics simple shear between two infinite parallel boundaries \cite{Itai}.  
	We applied a cyclic strain, whose maximum varied from 3\% to 10\%. In comparison, the yielding strain of this system is around~$3\%$~\cite{JieZheng}.  
	To study the effects of friction, we used: gear-shaped nylon particles (friction coefficient $\mu\to\infty$)~\cite{Corey}, a combination of  stainless-steel cylindrical rings
	and gear particles, photoelastic disks ($\mu\approx0.7$)~\cite{JiePEFriction}, ABS plastic cylindrical rings ($\mu\approx0.32$), stainless-steel cylindrical rings
	($\mu\approx0.3$), and Teflon-coated photoelastic disks ($\mu\approx0.15$).
	To minimize crystallization, we used a 50\%-50\% binary mixture of particles of size ratio 1:1.4. The diameter of the small particles was $1.0$~cm, except for gear particles, whose small particles had a pitch diameter of $1.6$~cm and a tooth height of $0.36$~cm. The total number of particles was about $2000$ in each system.  Each experimental run was started by depositing particles randomly inside the stadium and applying a cyclic shear until the system reached a steady state. 
	Below we show that fewer than 100 cycles suffice to de-correlate the system from its initial state. 
	Halting at every cycle the quasi-static process at maximum negative strain, we took a snapshot of the system, which allowed to monitor the static structure and the dynamics stroboscopicallly.  
	At steady state, the resulting packing fractions were $0.74\pm0.01$ for the gears, $0.81\pm0.01$ for the photoelastic particles, and $0.83\pm0.01$ for the Teflon-coated photoelastic, stainless steel, and plastic particles. These values are consistent with the expectation that in a sheared system at constant pressure, increasing the friction results in a decreasing packing fraction.
	More details on the experiment are given in the Supplementary Materials (SM)\footnote{\label{SuppMat}See Supplemental Materials [url] for experimental details, and various data for different systems; which includes~\cite{inertia_number,RafiSam2007}}.
	We show the results for a strain amplitude of 5\% and the results for other values are qualitatively the same. 
	For each given configuration of particles, we first identified the contacts and then constructed the cells, see Fig.~\ref{fig1}(b-c)~\cite{RafiTakashi2014,RafiTakashi2017}. Our results should be insensitive to the exact definition of cells~\cite{Tordesillas,Smart}. 
	Due to the finite resolution, we had to use a threshold to define inter-particle contacts. We have checked carefully that the results and conclusions presented here are not sensitive to this threshold, see SM~\cite{Note1} for details.

	\begin{figure}[tbh]
		\includegraphics{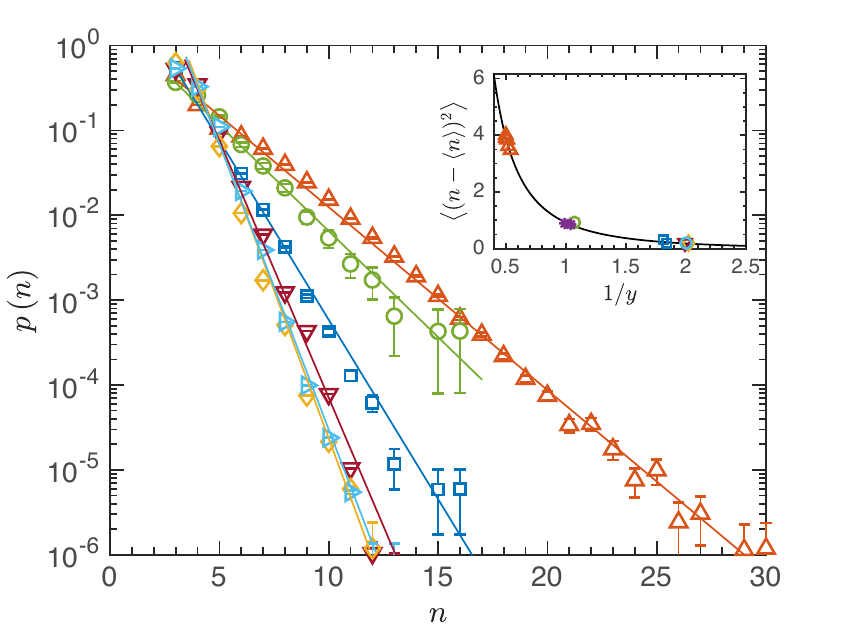}
		\caption{
			\label{fig2} 
			Cell Order Distributions of six different systems: Gears (triangles), gears/stainless steel cylindrical rings mixture (circles), photoelastic disks (squares), plastic cylindrical rings (inverted triangles), and Teflon-coated photoelastic disks (diamonds). 
			The solid lines are the exponential functions predicted by Eq.~(\ref{eq1}). Inset: 
			The variance of the COD for the different systems, with the same symbols as in the main panel and stars for photoelastic disks under pure shear. All measurements fall exactly on the analytical function predicted from the PDF in Eq.~(\ref{eq1}) The small spread of symbols within a given cluster of same-type particle represents variations of strain amplitudes from $3\%$ to $10\%$ and boundary pressure of $\pm30\%$.
		}
	\end{figure}
	The cell order $n$ is defined as the number of contacts that define the cell, see Fig.~\ref{fig1}(b) and (c) and the SM~\cite{Note1}. Figure~\ref{fig2} shows the cell order distribution (COD), $p(n)$, for six systems with different friction coefficients. The graph demonstrates that $p(n)$ is described very well by an exponential function, $p(n)\propto e^{-n/y}$. We find that the exponential form is independent of the friction coefficient, maximum strain, particles stiffness, and whether we applied a simple or pure cyclic shear~\cite{JieZheng}. As expected, the higher the friction the larger the occurrence frequency of high-order cells -~up to $n=30$ in the gear system~- enabling us to determine the distribution function to high accuracy. 
	Our CODs are somewhat different from those found in numerical studies of granular systems, in which the sample was gradually tilted~\cite{Smart} or isotropically compressed~\cite{RafiTakashi2014, RafiTakashi2017}, indicating that CODs: (i) are sensitive to driving protocol and (ii) differ between the quenched and quasi-static steady-states.
	
	We can understand this result in the context of the granular statistical mechanics approach~\cite{Rafi2003,Edwards1,Edwards2,Asenjo2014}.
	It is plausible that the structure self-organises via a competition between entropy, i.e.~increasing disorder with the largest possible number of cell configurations, and a constraint of mechanical equilibrium, which excludes unstable configurations.
	Expecting higher friction to enable more high-order cell configurations, thus increasing the importance of entropy, we neglect the stability constraint in our systems and use a maximum-entropy argument to derive the steady-state COD, $p(n)$. 
	We impose two constraints: that $p(n)$ is normalized and that it has a well-defined mean, $\langle n \rangle$, represented by two Lagrangian multipliers, $x$ and $-1/y$.
	Maximising the Gibbs entropy,\vspace*{-2mm}
	\begin{eqnarray}
	S = &-&\sum\limits_{n=3}^{\infty}p(n)\log{p(n)}+x\left[\sum\limits_{n=3}^{\infty}p(n)-1\right] \nonumber \\
	&-&\frac{1}{y}\left[\sum\limits_{n=3}^{\infty}np(n) - \langle n\rangle\right] \quad ,
	\label{EntMax}
	\end{eqnarray}
	
	\noindent
	yields\vspace*{-7mm}
	
	\begin{equation}
	p(n)=\frac{(1-e^{-1/y})}{e^{-3/y}} e^{-n/y} \quad ,
	\label{eq1}
	\end{equation}
	
	\noindent
	with $y= \left(\log\frac{\langle n\rangle-2}{\langle n\rangle-3}\right)^{-1}$ the typical decay of $p(n)$. 
	Using Euler's relation for planar graphs \cite{Euler} (see SM~\cite{Note1}), we also have 
	$y= \left(\log\frac{4}{6-\langle z\rangle}\right)^{-1}$, with $\langle z\rangle>2$ the mean coordination number.
	To test this result, we determined for each system the value of $\langle n \rangle$ and used it to get the corresponding value of $y$. In Fig.~\ref{fig2}, we include these theoretical predictions for $p(n)$ from Eq.~(\ref{eq1}) and find that the PDFs fall on the predicted curves almost perfectly - without any fitting parameter!
	
	The independence of the theoretical PDF of higher $n$ moments is non-trivial and reminiscent of the Boltzmann distribution, which depends only on the mean energy. In thermal statistical mechanics this is because the energy is defined up to an arbitrary constant~\cite{FeynmanBook2}. For the energy to be a proper extensive macro-quantity this constant must cancel on calculating its higher moments, which is only possible if the distribution is independent of the higher moments. In contrast, there is no physical reason why adding an arbitrary constant, $n\to n+n_0$, should not change the higher moments of $p(n)$. 
	To test whether or not higher moments need to be included via additional Lagrange multipliers, we compare  the experimentally computed variance of the COD, for different strain amplitudes, to that calculated from Eq.~(\ref{eq1}). The inset of Fig.~\ref{fig2} shows that the two coincide perfectly, establishing that the variance of the COD depends only on $\langle n \rangle$ and higher moments need not be included.
	
	\begin{figure}[tb]
		\includegraphics{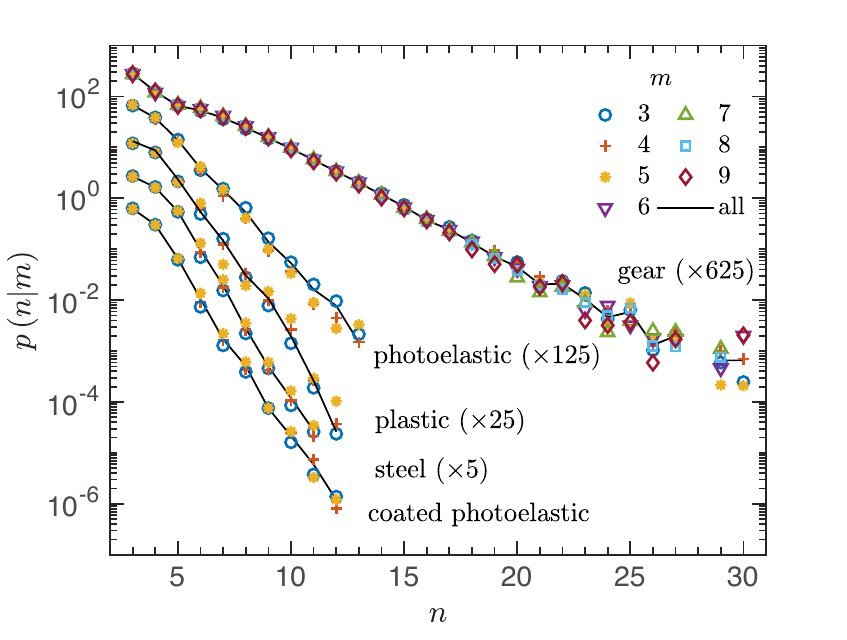}
		\caption{\label{fig3}
			The conditional CODs $p\left(n|m\right)$ for different systems. For clarity, data of stainless-steel particles, plastic particles, photoelastic particles and gears have been shifted vertically by factors of 5, 25, 125, and 625, respectively.}
	\end{figure}
	
	In this derivation we also neglect spatial cell-order correlations and to test this assumption we measured the conditional probabilities that a cell of order $n$ is neighbor of a cell of order $m$, $p(n|m)$. In Fig.~\ref{fig3}, we show these conditional PDFs for different values of $m$ for gear particles, photoelastic disks, plastic rings, stainless-steel rings, and coated photoelastic disks. For all systems, the PDFs collapse onto a master curve, establishing that $p\left(n|m\right)$ is independent of $m$ and that hence the cell orders of neighboring cells are not correlated. 
	This observation and the success of the maximum-entropy derivation support the view that local entropic effects play a key role in determining the cell structures.

	A way to characterize cells geometry is by the distribution of their internal angles, $\theta$. Figure~\ref{fig4} shows the conditional PDF of $\theta$, given the cell order, $q(\theta|n)$, for three representative friction systems: the gear (the highest friction, $\mu\to\infty$), photoelastic (intermediate friction, $\mu=0.7$), and Teflon-coated photoelastic systems (low friction, $\mu=0.15$). The PDFs are similar for $n\leq5$, mainly due to geometric constraints: when $n$ is small then, given the particle radii and their sequence around a cell, only $n-3$ internal angles are variable, limiting the explorable configuration space.
	In contrast, the PDF tails for $\theta> 180^o$ are increasingly suppressed with reducing friction. This corresponds to inhibiting too elongated cells~\cite{RafiTakashi2017}, which suppresses disorder and further supports our thesis that friction tunes the entropy-stability competition.
	
	\begin{figure}[tb]
		\centerline{\includegraphics{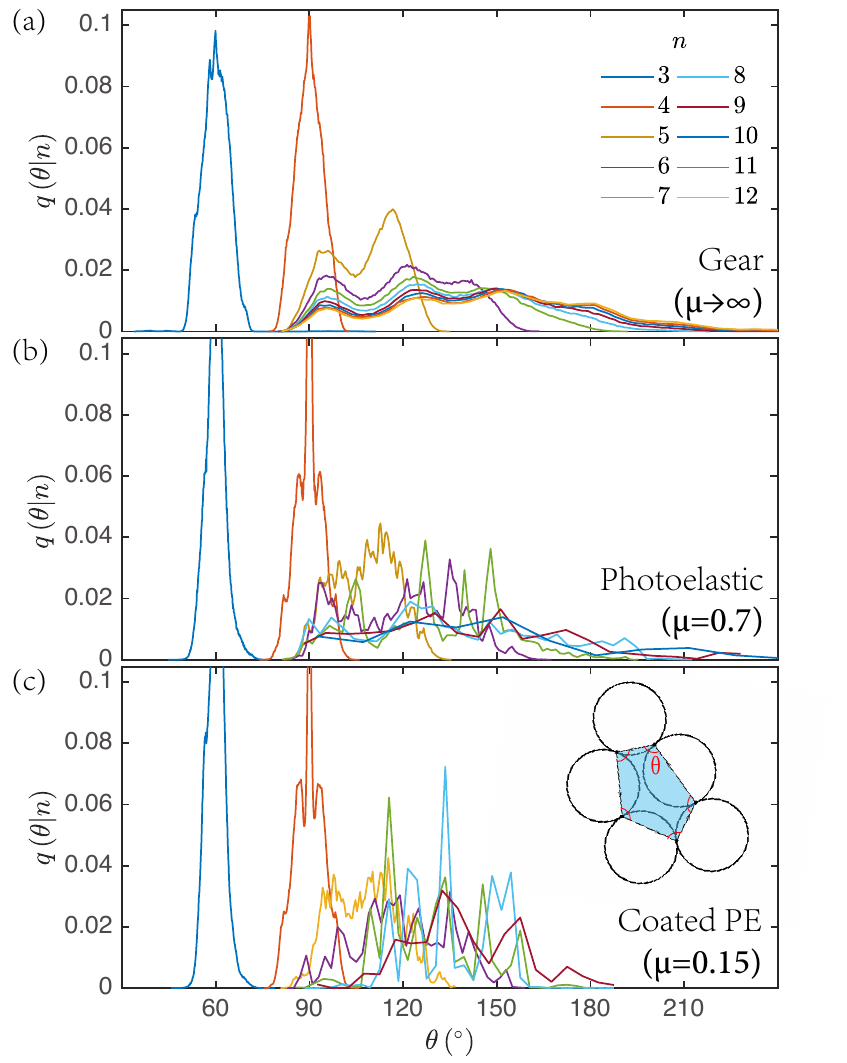}}
		\caption{\label{fig4}
			The conditional PDFs of the internal cell angles, defined in the inset, for: (a) gears; (b) photoelastic particles; Teflon-coated photoelastic particles.}
	\end{figure}
	
	\begin{figure}[htb]
		\includegraphics{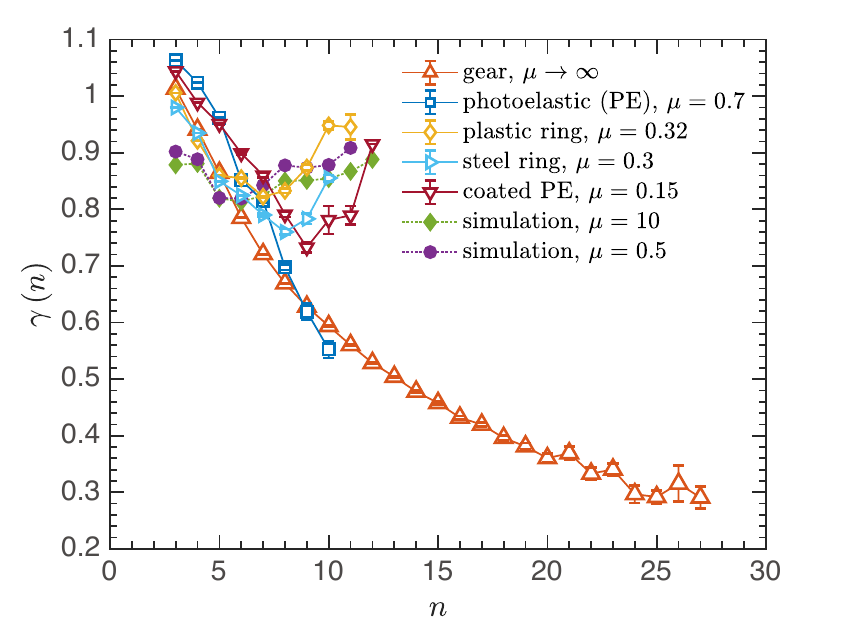}
		\caption{\label{Gamma}
			The ratio $\gamma =V_c(n)/V_{rp}(n)$ increases to $1$ as cells are more round and vice versa. Increasing entropy reduces $\gamma$, opposing the effect of mechanical stability. The monotonic decrease of $\gamma(n)$ at high friction indicates that entropy dominates over stability, while the up-turn of $\gamma(n)$ at low friction is evidence to the increasing restraining effect of stability. The latter behavior was also observed in~\cite{RafiTakashi2017} (included with permission), albeit under different dynamics.}
	\end{figure}
	
	This competition between the roughly exponential increase of the total number of cell configurations with $n$ and the restraining of too elongated cells also affects the distribution of cell aspect ratios. We studied this effect by investigating the ratio of the mean volume of $n$-cells to that of the equivalent regular $n$-polygon, $\gamma(n)\equiv V_{\rm c}(n)/V_{\rm rp}(n)$. The more severe the inhibition of elongated cells the closer is $\gamma$ to $1$. In other words, increasing entropy reduces $\gamma$ while the stability restrains this trend. 
	Figure~\ref{Gamma} demonstrates this competition and shows how friction modulates it: entropy dominates in high-friction systems, with $\gamma$ falling monotonically, while in low-friction systems the stability constraint eventually forces $\gamma$ up as $n$ increase. The value of $n$, when $\gamma$ starts increasing, depends on both friction and the driving protocol. For example, in the compression processes studied in \cite{RafiTakashi2017} (also shown in Fig. \ref{Gamma}), $\gamma$ eventually increases for both low and high friction.
	
	An important aspect of the dynamic structural self-organization is the lifetime of cells. Cells appear and disappear as contacts are made and broken~\cite{RafiTakashiGranMatt}. We define $S(t)$ - the probability that a cell existing at time $t_0$ neither merges with another cell nor splits until time $t_0+t$, with time measured in units of cycles. As shown in Fig.~\ref{fig5}, $S(t)$ decays sharply, a trend consistent with the assumption that merging and splitting of cells is a local, uncorrelated process. Using exponential fits for guidance, the relaxation times are short: $\tau \leq 0.84$, $\leq 1.43$ and $\leq 0.45$ for the gear, photoelastic and Teflon-coated photoelastic particles, respectively.
	The cell lifetimes in the Teflon-coated photoelastic system are significantly shorter than in the high-friction systems, consistent with the increased mechanical fragility as inter-particle friction is lowered.
	These results support our observation that the steady state is reached within at most a few dozen of cycles. 
	The decreasing lifetime with $n$ is due to a decreasing mechanical stability and is another fingerprint of the interplay between process-governed and friction-tuned entropy-stability competition. 
	Although the decay of $S(t)$ is faster for small $n$ in the photoelastic particles than in the gears, this trend is reversed for large $n$, consistent with the expected increased stability of high-order cells with friction. 
	Including in Fig.~\ref{fig5} the survival probabilities of rattlers in these systems, we observe that they remain disconnected from the network for longer than the lifetime of any cell order, suggesting that, to lowest order, their effect on the structural organization can be neglected in our dynamics. The rattlers survival time, while always longer than those of cells, also drops with decreasing friction indicating the rapid reconstruction and change of topology in low-friction systems.
	
	\begin{figure}[htb]
		\includegraphics{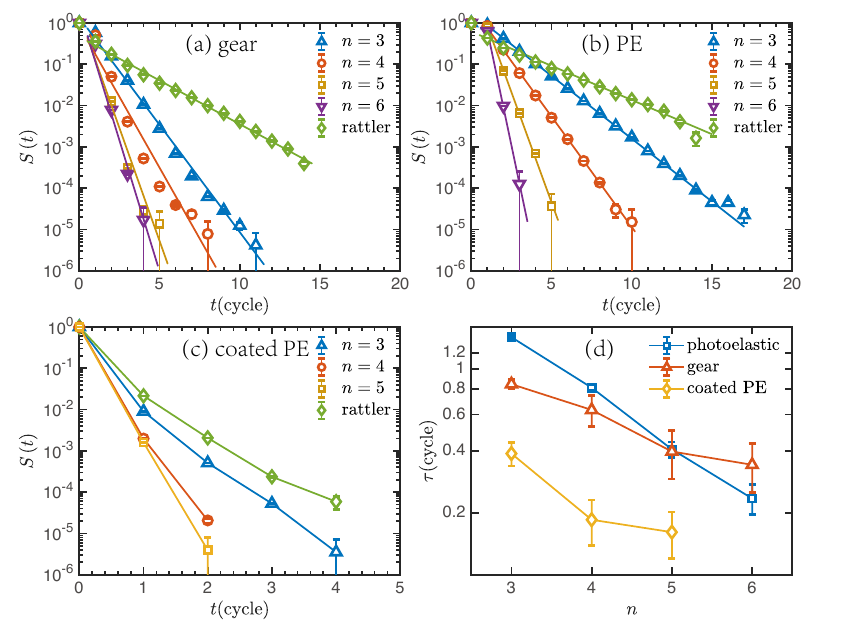}
		\caption{
			(a-c) The survival probability of cells of different orders, $n$, and rattlers, in the systems of (a) gears, (b) photoelastic particles, and (c) coated photoelastic particles. (d) The corresponding characteristic relaxation times, $\tau(n)$, estimated from an exponential fit of the data in the main panel.}
		\label{fig5}
	\end{figure}
	
	Our observations have another implication. A number of works in the literature model granular systems as glasses, especially in the quasi-static regime, where dynamic processes are slow. Yet, the short relaxation times we observe suggest that, in our process, the medium behaves rather as a liquid. Whether this conclusion extends more generally to quasi-static dynamics at lower shear rates and how it depends on the density, especially near the jamming point, remain open questions.
	Intriguingly, our high-friction systems contain a non-negligible fraction of particle strands with $z=2$ (see SM~\cite{Note1}). Similar strands, of dynamically cross-linked particles, have been observed in what is known as `empty liquids'~\cite{Sciortino,SciortinoExp,RuzickaExp,BiffiExp}. Whether this similarity provides a route to study empty liquids via macroscopic high-friction granular systems remains to be explored. 
	
	Our work shows that the value of the inter-particle friction coefficient does not affect the fundamental structural characteristics of quasi-static dynamic granular systems qualitatively. 
	Rather, it modulates the entropy-stability competition, expanding the cell configurations space as it increases. 
	This tunes the COD and affects the packing fraction and the cell survival time distributions. Finally, being able to understand the cell structural characteristics from entropy considerations supports the premise of Edwards and collaborators that granular systems can be described by entropy-based statistical mechanics~\cite{Rafi2003, Edwards1, Edwards2,EdwardsReview}.
	
	\bigskip
	
	\begin{acknowledgments}
		XLS and JZ thank I. Procaccia for valuable discussions. XLS and JZ acknowledge the student innovation center at Shanghai Jiao Tong
		University. WK is member of the Insitut Universitaire de France. This work is supported by the NSFC (No.11774221 and
		No.11974238).
	\end{acknowledgments}


%

\clearpage
\newpage

{\Large Supplementary Material:\\}

\renewcommand{\figurename}{Figure}
\renewcommand{\thefigure}{S\arabic{figure}}
\setcounter{figure}{0}

\title{Friction-controlled entropy-stability competition in granular systems: \\Supplementary Materials}

\author{Xulai Sun}
\affiliation{School of Physics and Astronomy, Shanghai Jiao Tong University, Shanghai 200240, China}

\author{Walter Kob}
\affiliation{School of Physics and Astronomy, Shanghai Jiao Tong University, Shanghai 200240, China}
\affiliation{Laboratoire Charles Coulomb, University of Montpellier and CNRS, F34095 Montpellier, France}

\author{Raphael Blumenfeld}
\affiliation{Gonville \& Caius College and Cavendish Laboratory, University of Cambridge, Cambridge CB2 1TA, UK}

\author{Hua Tong}
\affiliation{School of Physics and Astronomy, Shanghai Jiao Tong University, Shanghai 200240, China}

\author{Yujie Wang}
\affiliation{School of Physics and Astronomy, Shanghai Jiao Tong University, Shanghai 200240, China}

\author{Jie Zhang}
\email{jiezhang2012@sjtu.edu.cn}
\affiliation{School of Physics and Astronomy, Shanghai Jiao Tong University, Shanghai 200240, China}
\affiliation{Institute of Natural Sciences, Shanghai Jiao Tong University, Shanghai 200240, China}


\maketitle

In this Supplementary material we provide additional information on: details of the experimental setup; the procedure to identify contacts; effect of particle-support friction; the cell volume calculation; Euler relation; the effects of boundary conditions on the structure; 
the cell internal angle distributions and cell survival probabilities in the systems of plastic rings and stainless steel rings;
and the structural properties of the system. \\

{\bf 1. Experimental details}\\
During the experiment, particles were cyclically sheared within the stadium shear device driven by a stepping motor. In our system, we defined strain as the ratio between the displacement of lateral boundary and the width of the shear device, i.e.~the distance between the two lateral boundaries. 
The strain was varied from 3\% to 10\%. This maximum strain ensured that not too many particles would leave the camera window during a cycle, which could affect the statistics. 
The shearing was halted every cycle at maximum negative strain for two seconds and two snapshots were taken to improve the accuracy of the image analysis. 
The granular assembly was sheared by the shearing belt, which was kept under a constant confining pressure of $12.5$N/m. 
We estimate the inertial number of the particle motion at ${\cal{O}}\left(10^{-4}\right)$, allowing us to regard the process as quasi-static~\cite{inertia_number}. A snapshot of the system was taken every second to allow tracking of all the particle trajectories during each cycles.

Following the initial random placing of particles, several hundred shear cycles were applied; Fig.~6 of the main text shows that this is sufficiently long to decorrelate the system and to reach a steady state. Each data set presented in the main text and here was obtained in steady state from snapshots over approximately $4000$ cycles in the gear and mixed particles systems, $400$ cycles for the photoelastic and Teflon-coated photoelastic systems, 2000 cycles for the plastic particles system, and 1200 cycles for the stainless steel particles systems. The number of cycles in photoelastic and Teflon-coated photoelastic particles is smaller than the others because of a small, but finite, probability of the particles to buckle away from the plane, invalidating a large number of runs. 

In the mixed particle systems, the number of each kind of particles (large or small, gear or steel) were identical. At steady state, the resulting packing fractions were $0.74\pm0.01$ for the gears, $0.81\pm0.01$ for the photoelastic particles, and $0.83\pm0.01$ for the Teflon-coated photoelastic, stainless steel, and plastic particles. These values are consistent with the expectation that in a sheared system at constant pressure, increasing the friction results in a decreasing packing fraction.

\begin{figure}[tb]
	\includegraphics[width=0.95\columnwidth]{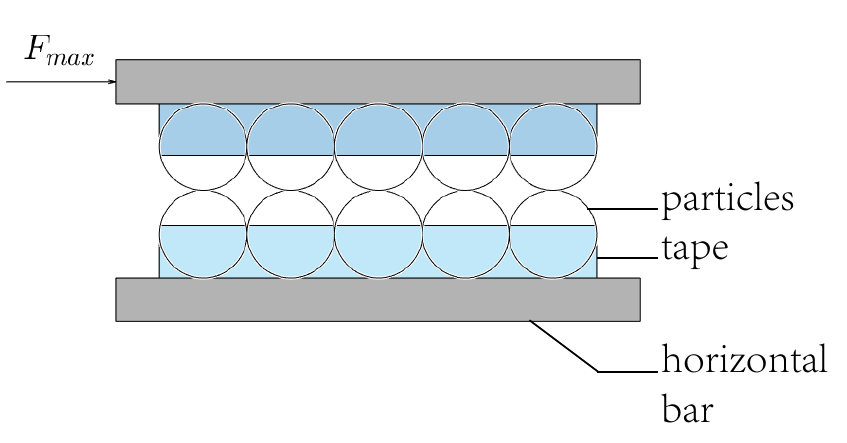}
	\caption{Sketch of the set-up for measuring the inter-particle friction coefficients.}
	\label{si_friction}
\end{figure}

Except for the gears, the inter-particle friction coefficients were measured as follows. Two identical arrays of same-size and same-material touching particles were aligned on flat bars and positioned vertically face to face, as shown in Fig.~\ref{si_friction}. An increasing horizontal force, measured by a commercial gauge, was applied to the top array until it started sliding when it reached $F_{max}$. The friction coefficient was determined as the ratio of $F_{max}$ to the total weight of the top array.\\

{\bf 2. Procedure to identify contacts}\\
We used the center-to-center distance as the main parameter to detect a contact between two neighboring particles. 
To correct for possible lens distortion effects, we tiled every image into a grid of squares $80$ pixels by $80$ pixels. 
Within each tile of each image, we computed the local pair correlation function and then averaged it over images of the same tile in different cycles to obtain the ensemble-averaged local pair correlation function. The first and third peak of this function were then used to determine the local radii of the small and large particles, respectively. 

\begin{figure}[tbh]
	\includegraphics{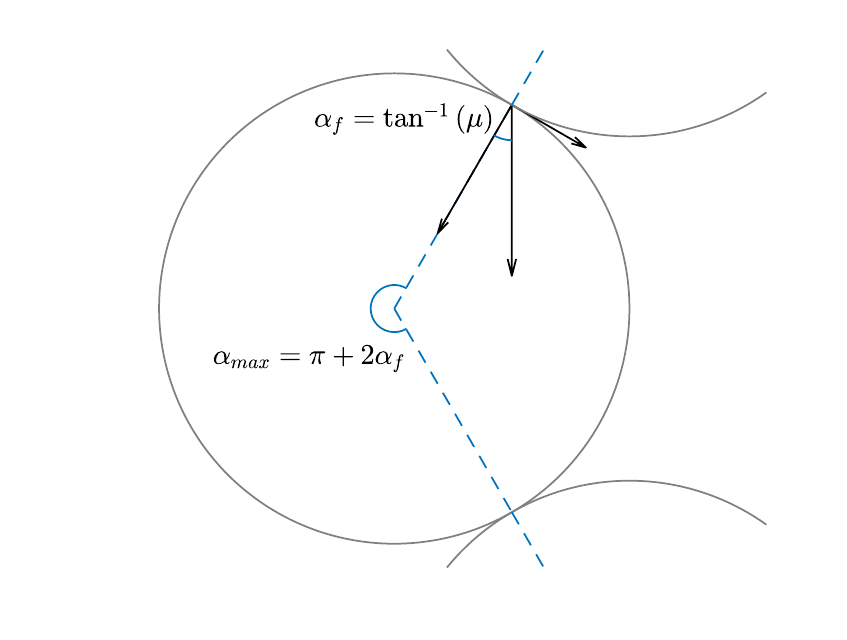}
	\caption{Sketch of a two-contact particle in a critical force-balance state.
	}
	\label{si_del}
\end{figure}

If the center-to-center distance between neighboring particles fell below the sum of their local radii plus a tolerance value, the particles were deemed to be in contact. A gear local radius is its pitch radius.
Once contacts are identified within one image, we checked that the contact positioning around each particle does not violate mechanical stability. For a given inter-particle friction coefficient, the friction angle at a contact between two particles is $\alpha_{f}=\tan^{-1}\left(\mu\right)$, which is defined as the angle between the contact force and its normal component, as shown in Fig.~\ref{si_del}. In order not to violate mechanical stability, the reflex angle specified between any two neighboring contact points on a given particle, as shown in Fig.~\ref{si_del}, should not exceed $\alpha_{max}=\pi+2\alpha_{f}$, otherwise at least one of the two neighboring contact point is a false contact, which leads to the direct identification of rattlers among particles of only two contact points. This process was repeated until no particles in the system violate mechanical stability. This check was particularly useful for particles with two or three contacts and allowed us to both eliminate all false contacts and remove potential rattlers.

\begin{figure}[tbh]
	\includegraphics{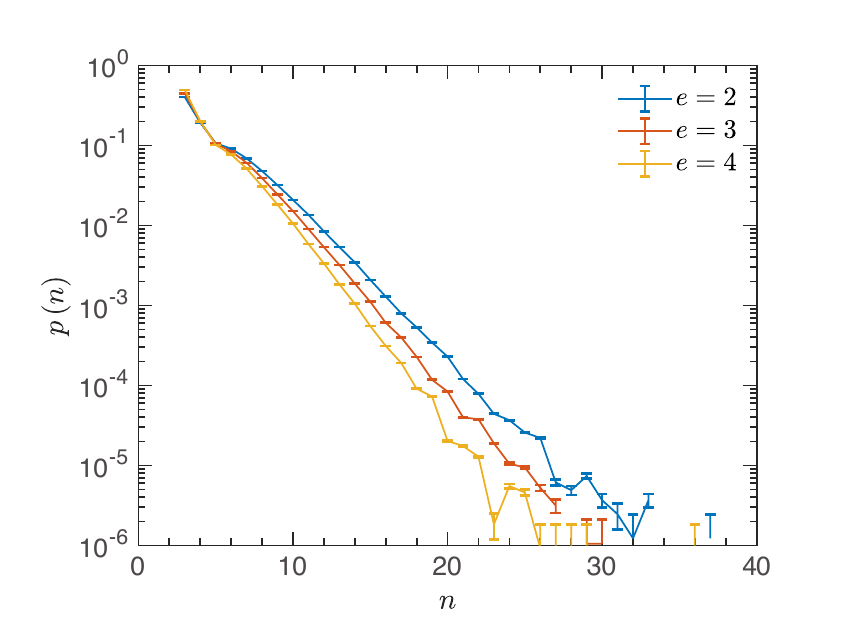}
	\caption{Effect of the finite resolution of contact determination on the cell statistics in the gear system. Cell order distributions calculated using different error tolerance values of 2, 3, and 4 pixels as specified in legend, show no qualitative difference, with the decay rate changed by at most 12\%.
	}
	\label{si_cod_error}
\end{figure}

At our image resolution, the pitch radii of the large (small) gears were $43$ ($34$) pixels and the radii of the large (small) photoelastic, Teflon-coated photoelastic, plastic, and stainless steel particles were $30$ ($22$) pixels. 
For the results presented in this paper we set the tolerance values of 2 pixels for photoelastic and steel particles, 2.5 pixels for plastic particles, and 3 pixels for gear particles. To test the effect of the finite resolution on contact determination, we varied the tolerance between 2 and 4 pixels and checked the effect on the cell order distributions. We found that the exponential form, Fig.~2 in the main text, was unaffected and that the decay rate changed by at most 12\%. As an example we show in Fig.~\ref{si_cod_error} the case of the gear system, demonstrating that the value of the threshold does not influence the functional form of the distribution. We point out that the mentioned 12\% change is an upper bound, at least for photoelastic and Teflon-coated photoelastic systems. For example, we used a series of snapshot from a pure shear experiment in \cite{JieZheng} and applied the above contact detection algorithm. We found that the number of identified contact points is only 4\% higher than the actual contacts identified using the combination of the above algorithm and the additional stress information from the force chain images.\\

{\bf 3. Effect of particle-base friction}\\
In principle, the friction forces between particles and the acrylic supporting base could affect the system dynamics and neglecting them requires justification. 
We measured the friction between each particle type and size and the base, $\mu_{b}$, and calculated the particle-base friction forces, $f_{b}=\mu_{b} m g$, with $m$ the particle's mass.
Under a confining pressure $P$, the inter-particle forces are typically $P D$, with $D$ the particle diameter. 
As can be seen from Table~\ref{si_table1}, for a typical pressure, $P=12.5$N/m, $f_b/PD$ is less than $0.06$ for all particles and hence the particle-base friction affects the dynamics only negligibly.

\begin{table}[tbh]
	\begin{tabular}{|l|l|c|c|c|c|}
		\hline
		\multicolumn{2}{|l|}{Particle type}    & $m$ (gram) & $D$ (cm) & $\mu_{b}$                         & $f_{b}/PD$ \\ \hline
		\multirow{2}{*}{gears}         & Large & 2.50   & 2.24            & \multirow{2}{*}{$\approx0.3$} & $0.026$                     \\ \cline{2-4} \cline{6-6} 
		& Small & 1.75   & 1.60            &                               & $0.026$                     \\ \hline
		\multirow{2}{*}{P.E. disks}         & Large & 0.99   & 1.40            & \multirow{2}{*}{$\approx0.3$} & $0.017$                     \\ \cline{2-4} \cline{6-6} 
		& Small & 0.50   & 1.00            &                               & $0.012$                     \\ \hline
		\multirow{2}{*}{plastic rings} & Large & 0.80   & 1.40            & \multirow{2}{*}{$\approx0.3$} & $0.013$                     \\ \cline{2-4} \cline{6-6} 
		& Small & 0.51   & 1.00            &                               & $0.012$                     \\ \hline
		\multirow{2}{*}{steel rings}   & Large & 3.40   & 1.40            & \multirow{2}{*}{$\approx0.3$} & $0.057$                     \\ \cline{2-4} \cline{6-6} 
		& Small & 2.27   & 1.00            &                               & $0.053$                     \\ \hline
		\multirow{2}{*}{Coated P.E. disks}         & Large & 1.13   & 1.40            & \multirow{2}{*}{$\approx0.3$} & $0.020$                     \\ \cline{2-4} \cline{6-6} 
		& Small & 0.58   & 1.00            &                               & $0.014$        \\ \hline
	\end{tabular}
	\caption{Particle parameters and the ratio of particle-base friction to inter-particle forces. Note that P.E. represents the abbreviation for `photoelastic'.}
	\label{si_table1}
\end{table}

\begin{figure}[tbh]
	\includegraphics{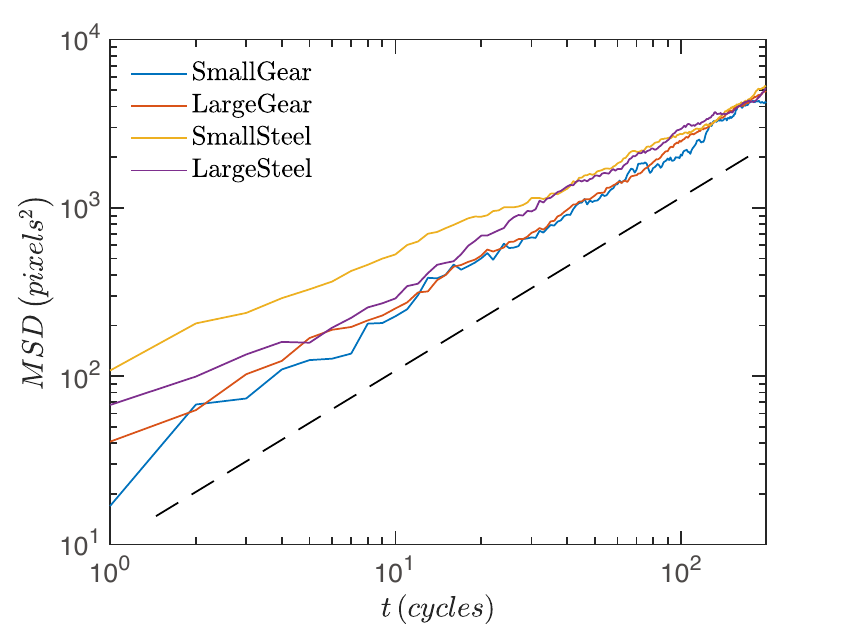}
	\caption{Mean squared displacement of the four different types of particles in the  mixed system.
		The dashed line is a power law with slope 1.0.
	}
	\label{si_msd}
\end{figure}

Since the particle-base friction depends on the particle masses, we had also to establish that it does not affect different-type particles sufficiently to take these into consideration. We therefore checked the mobility of the different-type particles in the system of gear/steel mixture.
The mean squared displacement (MSD) for the different kind of particles are shown in Fig.~\ref{si_msd}. While the small steel particles are more mobile than the large ones at short times, the MSDs of the two types, and in fact of all four types, become identical after 30-40 cycles. Moreover, the MSD of the lighter gear particles is in fact smaller at short times than that of the steel particles, indicating that particle masses hardly affect their mobility. Thus, the long-time dynamics is independent of particle type and the particles-base friction can be neglected.\\

{\bf 4. Effect of the stability-entropy
	competition on cell volume}\\
The structural properties of cells with a  given order, such as shape and volume distributions, are governed by the competition between the mechanical stability and entropy. The shape of a cell can be quantified not only by the internal angles distribution, as done in the main text, Fig.~4, but also by the ratio of cell volumes to that of the equivalent regular $n$-polygon, Fig.~5 in the main text. In this section we thus discuss how the volume of a cell depends on its cell order $n$.

To start, we note that cell edges can be defined in two ways, both of which give the same cell identity and order, but different volumes.

\begin{figure}[tbh]
	\includegraphics{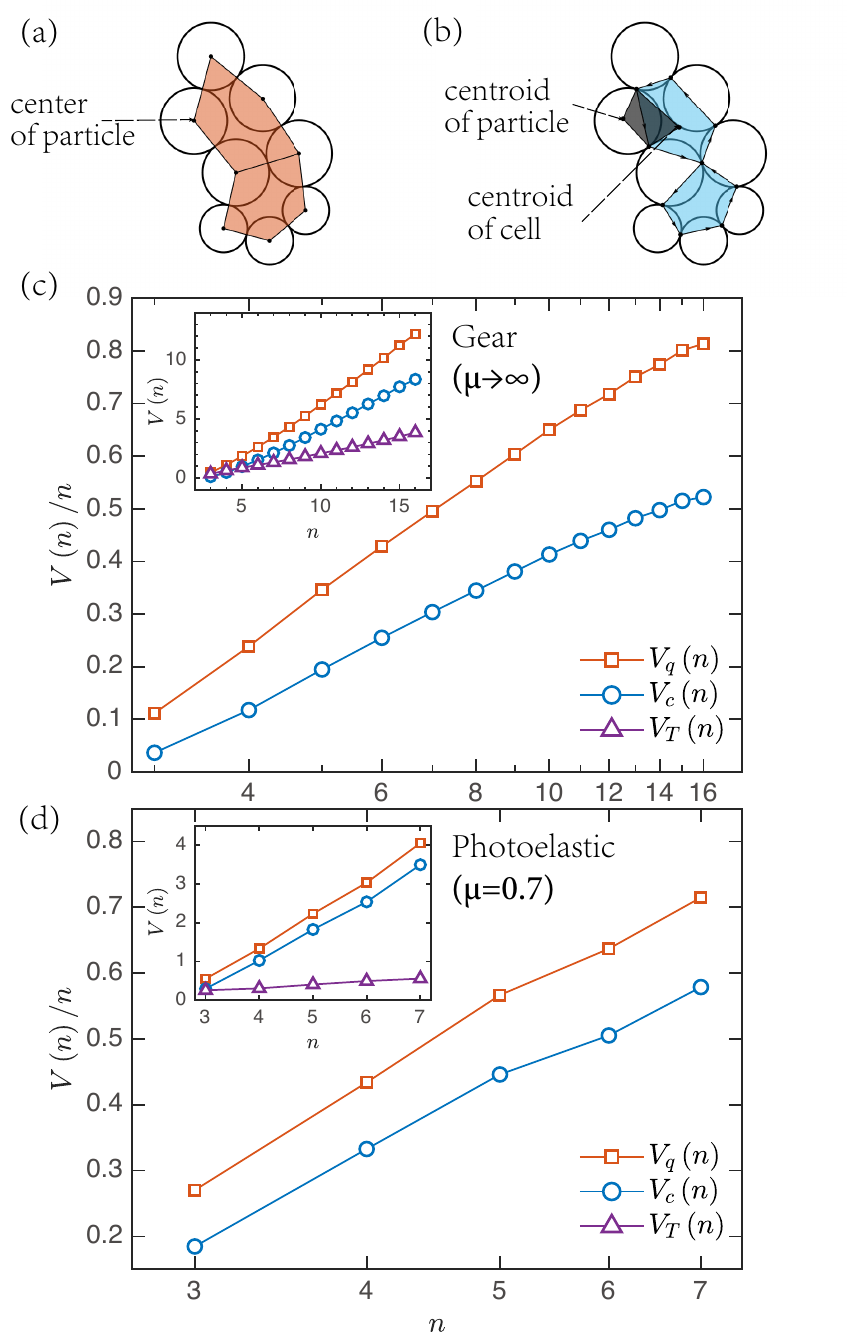}
	\caption{(a) Cells (shaded brown) defined by connecting particle centers. (b) Cells (shaded blue) defined by connecting particle contacts. The grey shaded region is an example of a quadron - the smallest volume element of the structure \cite{Rafi2003}.
		(c) Cell and quadron volumes for the gear system: The main panel displays the average volumes of cells per cell edge (circles) and of quadrons (square) as a function of $n$ with a logarithmic scale.  Inset: $n$-dependence (on a linear scale) of the volumes in the main panel, multiplied by $n$, and the the difference between them, $V_T(n)$  (triangles).
		(d) As in (c) for the photoelastic particles.}
	\label{si_fig1}
\end{figure}

Definition 1: The edges extend between neighboring particle centers around the smallest (aka irreducible) void loops, e.g. the brown regions in Fig.~\ref{si_fig1}(a). This definition is useful mainly for disks and particles with aspect ratio close to one.

Definition 2: The edges extend between contact points around such loops, e.g.~the blue regions in Fig.~\ref{si_fig1}(b). This definition yields smaller cells and it has two main advantages: it is not sensitive to the particle shape and it describes directly the contact network. Since the latter underlies stress transmission during the dynamics of granular materials, as well as in the static states of granular materials, definition 2 has been shown to be useful for the analysis of granular structures~\cite{RafiTakashi2017,Rafi2003} in that it allows one to define the smallest possible volume element in granular systems - the `quadrons'~\cite{Rafi2003}. A quadron is based on the contact network and both a particle and one of its neighboring cells contribute to its volume.
To construct a quadron (see also Fig.~\ref{si_fig1}(b)): (i) determine the centroid of a particle as the mean position of its contacts, $\vec{\rho}_g$; (ii) determine the centroid of one of its neighboring cells as the mean position of the contacts around it, $\vec{\rho}_c$; (iii) draw a vector between the two contacts of the particle that are shared with this cell, $\vec{r}_{cg}$; (iv) the quadron is the quadrilateral, whose diagonals are $\vec{r}_{cg}$ and $\vec{\rho}_c - \vec{\rho}_g$. An example of a quadron, shaded grey, is shown in Fig.~\ref{si_fig1}(b). 
The quadrons tessellate the system perfectly, except for wildly non-convex particles, in which case a small percentage of them may overlap~\cite{RafiSam2007}.

To analyze the cell volumes, we define the following quantities:
$V_c$, the volume of a specific cell as given by definition 2; 
$V_q$, the sum over the volumes of the quadrons `belonging' to a specific cell; 
$V_T$, the sum over the volumes of the triangles, each of which is defined by the centroid of a particle around the cell and the two contact points that the particles shares with it, or in other words, the part of the quadron within the particle (see Fig.~\ref{si_fig1}(b). 
By these definitions, $V_q=V_c+V_T$ for each cell. 

To test the dependence of the volumes on the cell order $n$, we plot the cell volume {\it per contact}, $V_c/n$, and the mean quadron volumes, $V_q/n$ against $n$ for the gear (Fig.~\ref{si_fig1}(c)) and photoelastic (Fig.~\ref{si_fig1}(d)) systems. 
In these plots, volumes are measured in units of $\pi\bar{R}^2$, with $\bar{R}\equiv\left(R_s+R_l\right)/2$, with $R_s$ and $R_l$ the radii of the small and large particles, respectively.
The mean of $V_T$ increases linearly with $n$, see insets, which is to be expected with this distribution of particle sizes.

\begin{figure}[tb]
	\includegraphics[width=0.95\columnwidth]{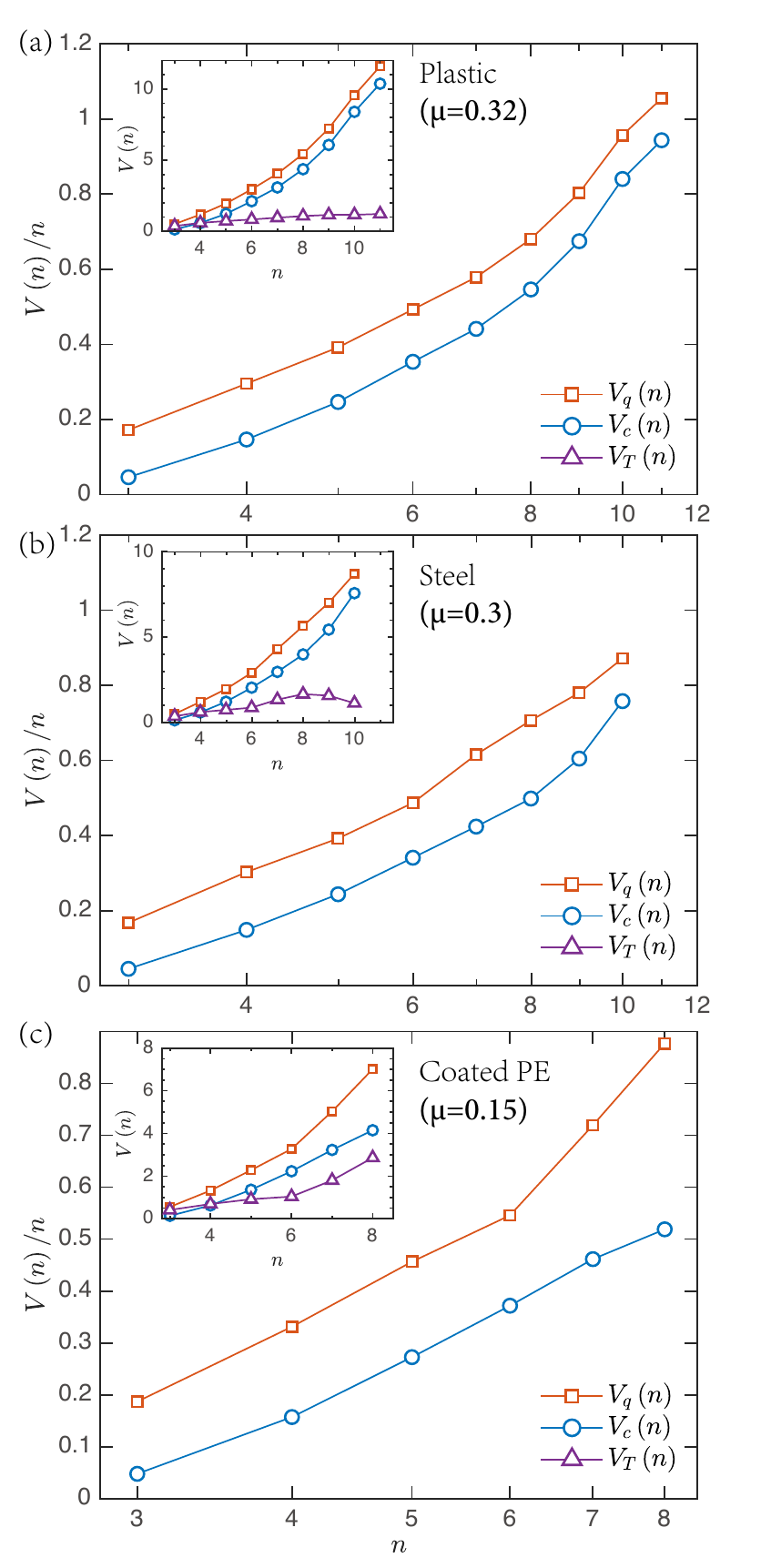}
	\caption{Cell and quadron volumes for the low-friction systems. The main panel displays the average volumes of cells per cell edge (circles) and of quadrons (squares) as a function of $n$ with a logarithmic scale.  Inset: $n$-dependence (on a linear scale) of the volumes in the main panel, multiplied by $n$, and the difference between them, $V_T(n)$  (triangles).
		(a) Plastic particles. (b) Steel particles. (c) Teflon-coated photoelastic particles.
	}
	\label{si_fig1_1}
\end{figure}

For these high friction systems an apparently reasonable fit for $V_c(n)$ is
\begin{equation}
V_c(n) = n[A\ln(n)-B] \quad ,
\label{eq_si1}
\end{equation}
with $A$ and $B$ constants, although we know of no theoretical model that predicts this form. The trend in high-friction systems shown in Figs.~\ref{si_fig1} can be understood as follows: At low values of $n$, geometry and mechanical stability constrain the mean volume of cells to approximate the equivalent regular polygons, 
$V_c(n)\approx V_{rp}(n)=n\bar{R}^2\cos^3(\pi/n)/\sin(\pi/n)$, albeit with some stretching along the shear-driven tilted principal stress. As $n$ increases, more elongated cells can be realized, which suppresses the mean volume increase rate. 

The increase in the fraction of elongated cells at a given cell order $n$ with inter-particle friction leads to a decrease in the average cell volume, an effect that is seen clearly in Fig.~S4: the higher the friction the slower the increase of the average cell volume with $n$. For example, at $n=7$ the value of $v_c/n$ is just under $0.6$ for the photoelastic system ($\mu = 0.7$) while it is below $0.4$ for the gears ($\mu\to\infty$).

\begin{figure}[tb]
	\includegraphics{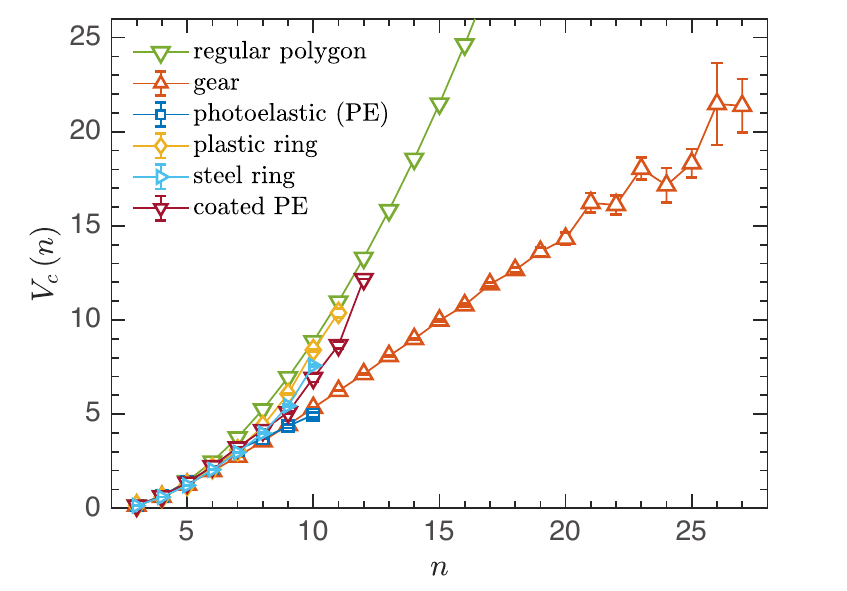}
	\caption{
		The $n$-dependence of the mean cell volume for the various experimental systems and of the volume of the  regular $n$-polygon. The volumes are measured in units of $\pi \bar{R}^2$. 
	} 
	\label{sGamma}
\end{figure}

In Fig. \ref{sGamma}, we plot $V_c(n)$ for all the systems considered. As expected, reducing the intergranular friction steepens the curves at large $n$ and they tend toward the limit of the regular polygon, included in the graph as well. While $V_{rp}\sim n^2$ for large $n$, we observe that, intriguingly, $V_c$ increases almost linearly with $n$ for high-friction systems. This is again evidence for the ubiquity of elongated cells in these systems, which are suppressed by the mechanical stability constraint in low-friction systems. This rapid increase of the configurations space in high-friction systems, in spite of the stability constraint, is the reason for the domination of entropy in determining the cell structures distribution.\\ 

{\bf 5. Euler relation and boundary corrections}\\
To check for potential influence of the boundary on our results, we first use Euler's relation to establish a relation between the particles mean coordination number, 
$\langle z \rangle$, and the mean cell order, $\langle n \rangle$~\cite{RafiTakashi2014},

\begin{equation}
\langle n\rangle = \frac{2\langle z \rangle }{\langle z \rangle -2}+O\left(N^{-1/2}\right) \quad .
\label{si_eq1}
\end{equation}

\begin{figure}[hbt]
	\begin{center}
		\includegraphics{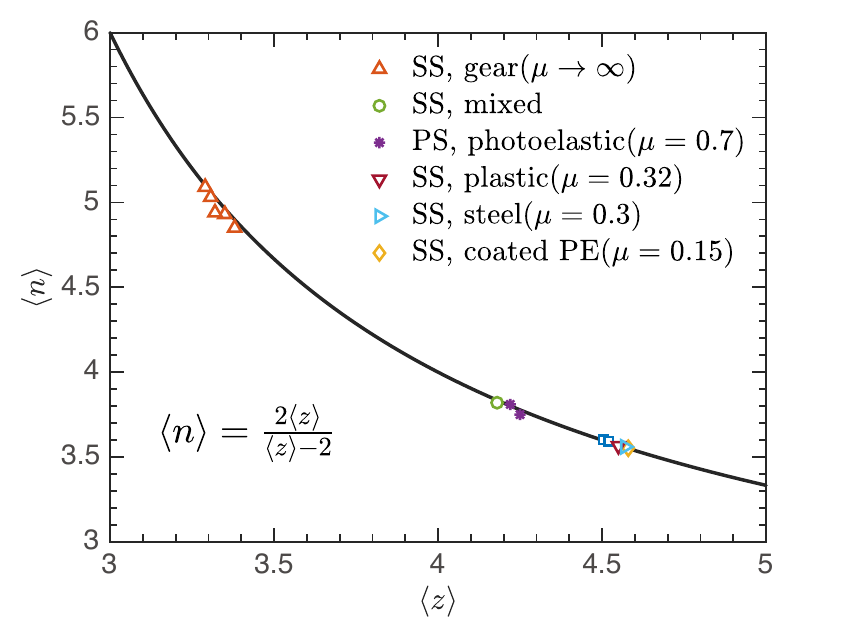}
		\caption{\label{si_fig2} 
			Check of the boundary correction, using Euler's relation, Eq.~(\ref{si_eq1}). The symbols represent measurements of $\langle z \rangle$ and $\langle n \rangle$, corresponding to different systems (see legend) in different conditions: strain amplitudes, etc. The solid line is Euler's relation for $N\to\infty$, in which limit the boundary correction vanishes.
			We observe that the experimental data fall nicely on top of the theoretical limit, establishing that the boundary effects are negligible. The small spread of symbols within a given cluster of same-type particle represents variations of strain amplitudes from $3\%$ to $10\%$ and boundary pressure of $\pm30\%$.}
	\end{center}
\end{figure}

\noindent
Here $N$ is the total number of particles and the second term on the right hand side represents boundary corrections. 
We then plot in Fig.~\ref{si_fig2} relation (\ref{si_eq1}) for $N\to\infty$ (solid line) and our measured values of $\langle z \rangle$ and $\langle n \rangle$ (symbols). 
Given sufficient statistics, which we have, any deviation between the experimental data and the curve would be evidence for existing finite boundary corrections.
The experimental data falls squarely on the theoretical curve, showing that the boundary corrections to our observations are negligibly small. \\

\begin{figure}[bht]
	\includegraphics{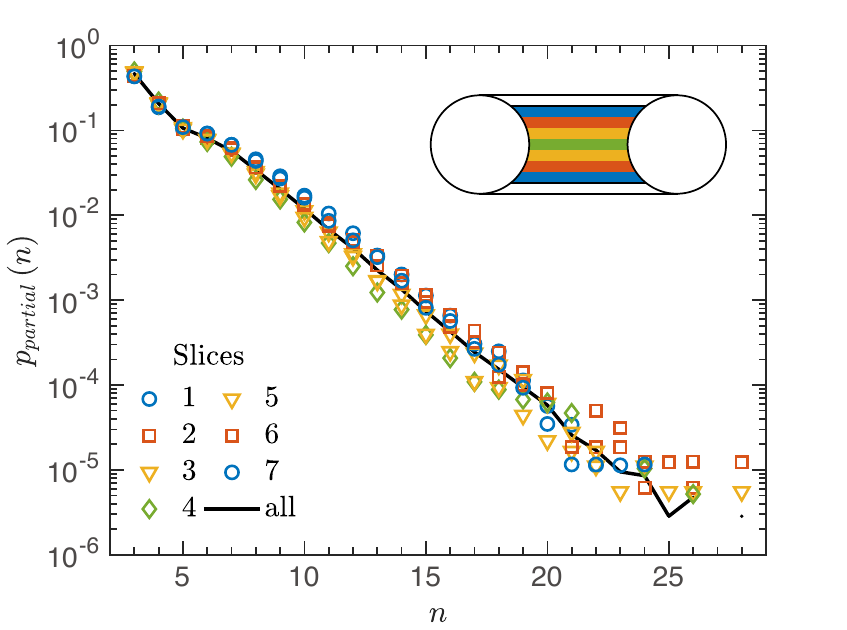}
	\caption{\label{sfig3}
		Cell order probability density functions (PDFs) for different slices of the gear system at a strain amplitude of $5\%$. The symbol colors correspond to the slice colors in the inset.
		No significant dependence of the PDFs on their position is observed. The solid black line is the average PDF of the system.} 
\end{figure}

{\bf 6. Homogeneity of the system}\\
In a setup for simple shear there is a danger of a density gradient developing between the boundaries. To check the structure for statistical homogeneity, we divided the system into nine slices parallel to the shearing boundaries, as shown in the inset of Fig.~\ref{sfig3}, and calculated the COD in each slice. Plotting all the PDFs in Fig.~\ref{sfig3}, we observe that these are all exponential for $n>5$, $p(n)\sim e^{-n/y}$, with $y$ depending very weakly on the slice's position. To a good accuracy, we can thus regard the PDFs as collapsing onto a master curve. The noise for very large values of $n$ we attribute to the decreasing statistics in this range of $n$. 
This collapse is consistent with the system being statistically homogeneous.\\

\begin{figure}[htb!]
	\includegraphics{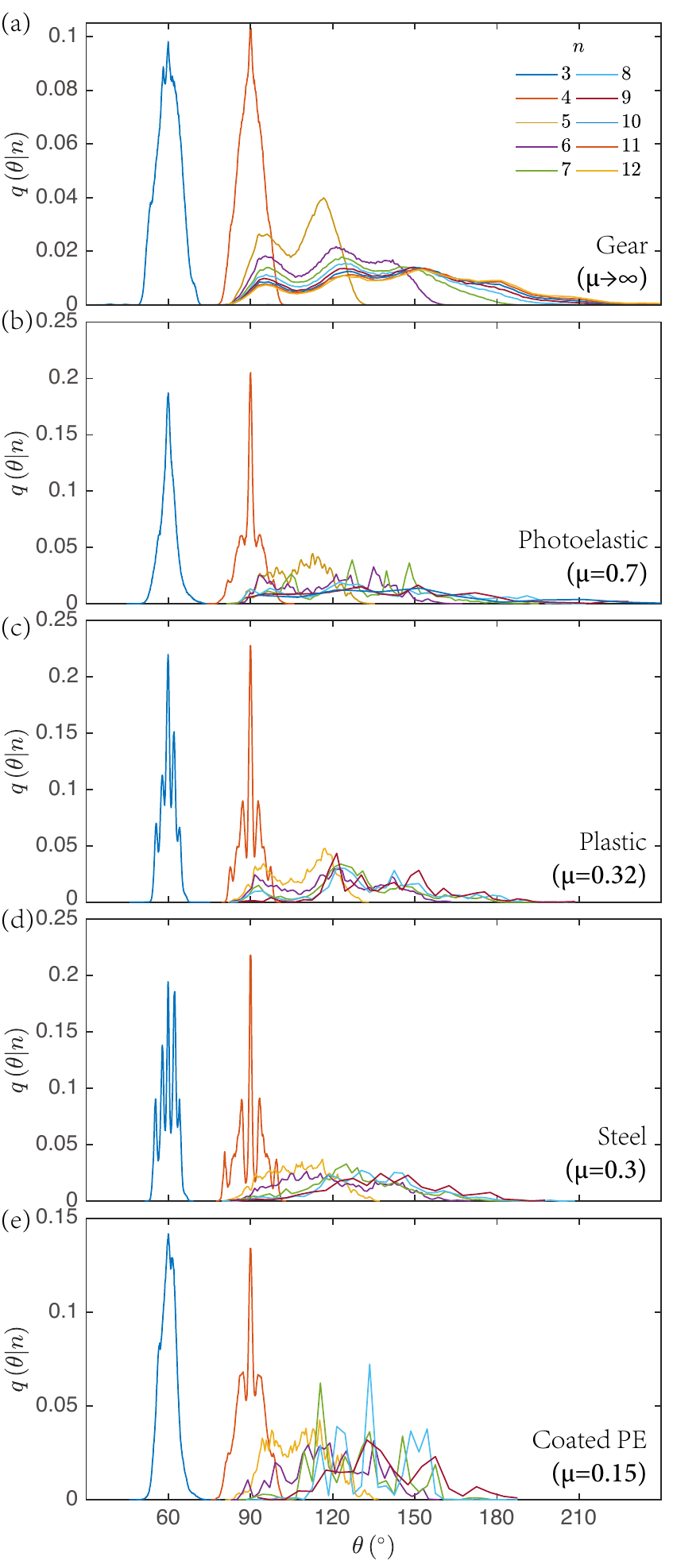}
	\caption{\label{sfig6} Distributions of the internal angle for different cell orders $n$ for various systems. (a) Gear system, (b) Photoelastic system, (c) Plastic particle system, (d) Steel particle system, (e) Teflon-coated photoelastic system.}
\end{figure}

\begin{figure}[tb]
	\begin{center}
		\includegraphics{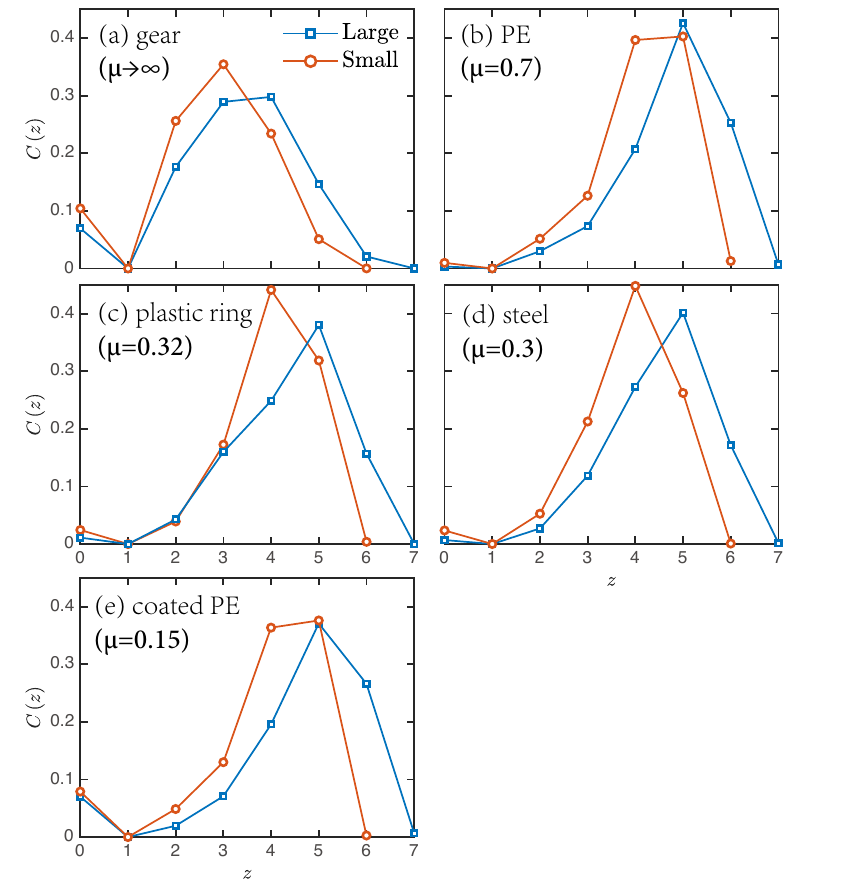}
		\caption{\label{sfig4} Distribution of the coordination numbers for the small and large particles. (a) gear, (b) photoelastic (PE) particles, (c) plastic ring systems, (d) steel particles, and (e) Teflon-coated photoelastic particles. The shear strain is 5\%.
		}
	\end{center}
\end{figure}

\begin{figure}[tbh]
	\includegraphics{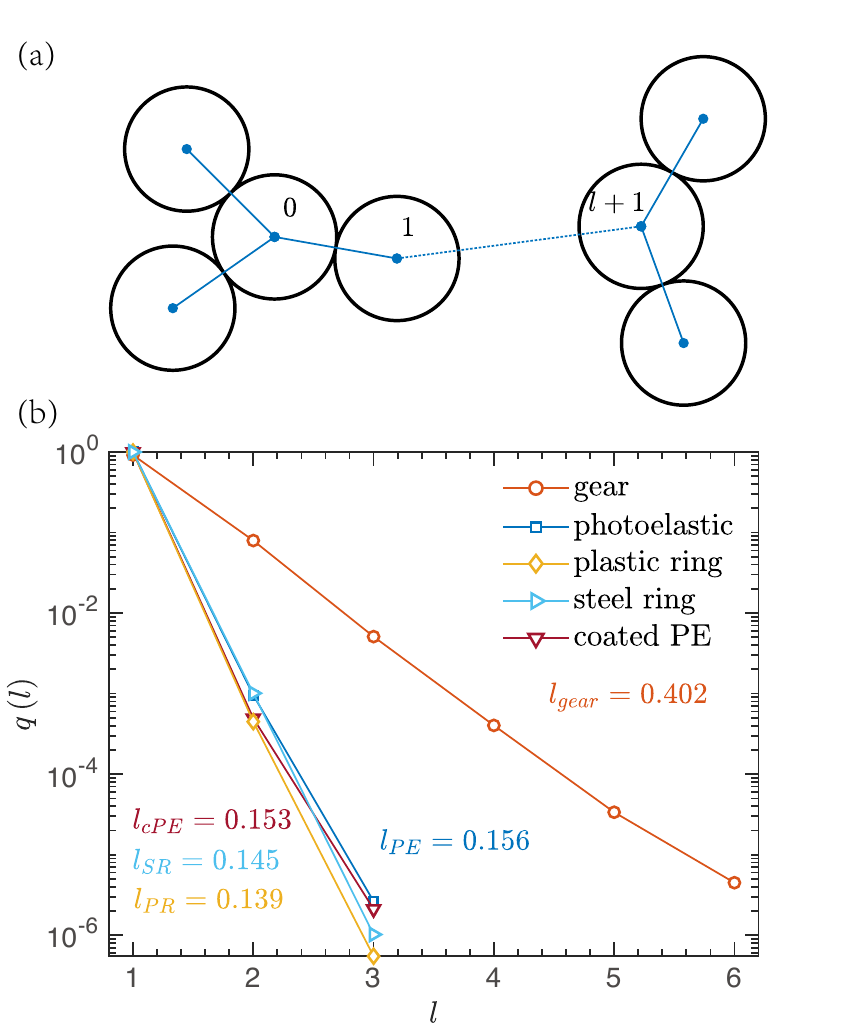}
	\caption{\label{sfig5} (a) Sketch of a strand of length $l$.
		(b) The strand length PDFs for the different systems. The strain amplitude is $5$\%.
	}
\end{figure}

{\bf 7. Internal angle distributions}\\
Similar to the case of Teflon coated photoelastic disks show in the main text, Fig.~4, the internal angle distributions in the other two low-friction systems, plastic rings ($\mu=0.32$) and stainless steel rings ($\mu=0.3$), are qualitatively similar (see Fig.~\ref{sfig6}). As is the case for the low-friction systems discussed in the main text, the PDF tail at angles larger than 180$^\circ$ is strongly suppressed due to the reduced mechanical stability of such cells. This is consistent with the plot in Fig.~5 in the main text, which shows that large cells in low-friction systems are more  rounded.\\

{\bf 8. Cells structural properties}\\
A key characteristic of the local structure is the distribution of the particles' coordination numbers, $C(z)$. Since these distributions are expected to depend on the particle sizes, we show in Fig.~\ref{sfig4} the $C(z)$ for the large and small particles separately. 

\begin{figure}[tbh]
	\includegraphics{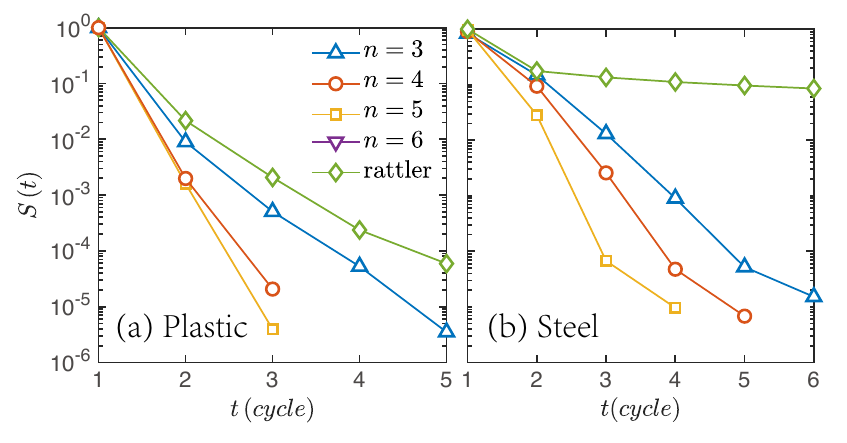}
	\caption{\label{sfig6_bis} 
		The survival probability of cells of different orders $n$ and rattlers in (a) Plastic particle system and (b) Steel particle system.}
\end{figure}

In the gear system the mean coordination numbers for the small and large particles are $z_s=3.09\pm0.07$ and $z_l=3.51\pm0.09$, respectively, Fig.~\ref{sfig4}(a). These values are significantly lower than the densest state, $\langle z\rangle = 6$, and are close to the marginally rigid limit of an infinite system, which indicates a very dilute structure. Figure~\ref{sfig4}(a) also shows that for about 10\% of the particles $z\leq1$ - these are rattlers that do not transmit forces.
The corresponding mean coordination numbers for the photoelastic particles are $z_s=4.17\pm0.03$ and $z_l=4.79\pm0.03$, respectively. These systems are denser than the gear systems, which is to be expected because of the relatively lower inter-particle friction. 
Lowering the friction further leads to smaller cells and increases the density and the mean coordination numbers: For the plastic particles we find $z_s=4.20\pm0.02$ and $z_l=4.79\pm0.02$, for the steel particles we have $z_s=4.21\pm0.03$ and $z_l=4.78\pm0.03$, and for the Teflon-coated photoelastic particles we have $z_s=4.17\pm0.03$ and $z_l=4.87\pm0.03$.

The fast structural relaxation of the system discussed in the main text suggests that these systems are liquid-like rather than glassy and because of the low density they resemble an empty liquids~\cite{Sciortino,SciortinoExp,RuzickaExp,BiffiExp}. A key feature of empty liquids is their internal structure, which consists of strands of coordination number $z=2$, whose lengths have been found to be given by an exponential distribution.

To check for this feature in our systems, we have calculated $q(l)$, the  probability that a strand has exactly length $l$, as sketched in Fig.~\ref{sfig5}(a). In Fig.~\ref{sfig5}(b) we show that, as in empty liquids, $q(l)$ is an exponential: $q(l)\propto \exp(-l/l_0)$. For all our systems we find that the decay length is small, $l_0=0.402$, $0.156$, $0.153$, $0.145$ and $0.139$ for our gear, photoelastic, Teflon-coated photoelastic, steel and plastic systems, respectively, thus much shorter than the values found in empty liquids strands~\cite{Sciortino,SciortinoExp,RuzickaExp,BiffiExp}. 
However, since $l_0$ depends on the details of the experimental setup, such as particle-base and interpaticle frictions, confining pressure, and possibly others, we cannot rule out the possibility that longer strands and lower densities may be achievable in other setups, which could then be used as convenient macroscopic models for empty liquids.\\

{\bf 9. Cell survival probabilities of plastic ring and stainless steel ring systems}\\
Here, we show that the survival probabilities of cell orders follow the same pattern as for the systems of gear, photoelastic, and coated photoelastic particle systems, discussed in the main text. In Fig.~\ref{sfig6_bis} we show the probability that a cell present at time $t=0$ has neither split nor merged with other cells at time $t$. These probabilities in the plastic and stainless-steel ring systems are qualitatively similar to the case of the Teflon-coated photoelastic particles shown in Fig.~6 in the main text. Note that the lifetime of rattlers in the steel particle system is significantly larger than that in plastic particle systems, indicating that an increase in the friction between the particles and the supporting base may slow down the integration of rattlers into the force carrying network.
\\

\vfill

	\end{document}